\documentclass[reprint,amsmath,amssymb,aps,pra,letter,showpacs]{revtex4-1}

\usepackage{graphicx}
\usepackage{dcolumn}
\usepackage{bm}
\usepackage{float}

\begin{document}


\title{Second Plasmon and Collective Modes in Binary Coulomb Systems}

\author{G. J. Kalman$^1$, Z. Donk\'o$^{1,2}$, P. Hartmann$^{1,2}$, K. I. Golden$^3$}
\affiliation{$^1$Department of Physics, Boston College, Chestnut Hill, MA 20467, USA}
\affiliation{$^2$Institute for Solid State Physics and Optics,
Wigner Research Centre for Physics,\\
Hungarian Academy of Sciences, H-1121 Budapest, Konkoly-Thege Mikl\'os str. 29-33, Hungary}
\affiliation{$^3$Department of Mathematics and Statistics and Department of Physics, University of Vermont, Burlington, VT 05405-1455, USA}

\date{\today}

\begin{abstract}
In a system consisting of two different charged species we identify the excitation of a second, low frequency plasmon. At strong coupling the doublet of high frequency (first) and low frequency (second) plasmons replaces the single plasmon excitation that prevails at weak coupling. We observe the formation of the second plasmon from the acoustic Goldstone type mode associated with short range interaction as the range is extended to infinity. 
\end{abstract}

\pacs{52.27.Gr, 52.27.Cm, 52.27.Lw, 52.25.Mq, 52.35.Fp, 52.35.Lv, 52.65.Yy, 71.45.Gm}

\maketitle

The existence of plasmons in many body systems interacting through a Coulomb potential (plasmas, electron gases, etc.) with a characteristic oscillation frequency, the plasma frequency
\begin{equation}
\omega_0 = \sqrt{4 \pi Z^2 e^2 n / m}
\end{equation}
(with the symbols having their usual meaning) has been known for a long time \cite{1}. The identification of this phenomenon as a collective excitation -- in fact, the very  introduction of the idea of collective excitations and the notion  of collective coordinates -- is due to the pioneering series of works by Bohm, Gross and Pines \cite{2,*3,4,5}. It was also Bohm and Gross \cite{2,*3} (BG) who determined the eponymous $k$-dependent positive dispersion of the plasmon, caused by the random motion of the particles. Soon, however, it became clear that the BG dispersion and the underlying theoretical approach (which later was reformulated in many differ guises \cite{4,26,*24,*17} and has commonly become known as the Random Phase Approximation (RPA)) are appropriate for weak coupling only. The coupling strength is conveniently defined as the ratio of the potential energy of the particles to their kinetic energy. The appropriate parameters  that characterize the coupling strength for classical systems are $\Gamma=Z^2e^2/ak_\text{B}T$ and for quantum systems $r_s=a/a_B$ ($a$ is the Wigner-Seitz radius, $a_B$ the Bohr radius and $T$ the temperature). Motivated by the case of the electron gas in metals where the condition $r_s<1$ is mildly violated, it was Singwi and collaborators \cite{7} who have made the first serious attempts to study  the effect of strong coupling on  the properties of the plasmon. However, the first systematic and reliable analysis of this problem, primarily through molecular dynamics (MD) computer simulation was done by Hansen and collaborators \cite{8,*9,*10,27}; in particular Hansen \cite{27} verified the change of the BG behavior to a negative dispersion, the hallmark phenomenon of strong coupling, which was predicted and investigated by a number of workers \cite{12,*13,14a,*14b} around the same time. A different theoretical approach, the Quasilocalized Charge Approximation (QLCA), geared for the study of strongly coupled Coulomb systems was introduced by Kalman and Golden \cite{21,22} and combined with advanced MD computer simulations has led to a thorough investigation of the plasmon dispersion. Experimentally, the plasmon dispersion of the electron gas has been mapped directly and indirectly  in various condensed matter situations at low or moderate coupling values; with the advent of complex (dusty) plasma experiments, the  way to directly observing strongly coupled plasmon behavior in the laboratory has opened up.
  
Looking at the problem from a more general point of view, we focus first on a system governed by a short range interaction (e.g.  by a  Yukawa potential, $\varphi(r) \propto {\rm e}^{-\kappa r}/r$). Such a system exhibits  three $\omega(k \rightarrow 0) \sim k$  acoustic Goldstone type excitations \cite{28}, one of which  is a longitudinal mode.  This, however, is not the case for a plasma with long range (i. e. $\kappa=0$) Coulomb interaction. The fundamental work of Anderson \cite{15} has shown that this zero mass Goldstone boson acquires a mass to transform itself into the finite mass longitudinal plasmon with $\omega(k \rightarrow 0) = \omega_p$. It has also been demonstrated by Lange \cite{16} that the argument that associates the generation of a Goldstone boson with a broken symmetry fails for long range interaction. Moreover, protected by the Kohn Sum Rule \cite{18}, the plasmon is an extremely robust excitation, unaffected by correlations, i.e. $\Gamma$ and $r_s$ independent.

The question we address now in this Letter is what happens then in a (three dimensional) binary Coulomb system, composed of two species of different masses and charges? Choosing the Yukawa system again as a starting paradigm, we observe that the system now exhibits, in addition to the longitudinal acoustic mode, a longitudinal optic mode, which at $k=0$ has frequency $\omega_\ast$ and is degenerate with its two transverse counterparts (cf. the corresponding discussion on the 2D system in \cite{20}). Following Anderson's argument, we now expect that with the Coulomb interaction switched on, the acoustic excitation acquires a mass, i.e. develops a finite frequency, and becomes a new excitation $\omega(k \rightarrow 0) = \omega_-$ which we refer to as the low frequency second plasmon. It is less obvious what happens to the gapped  longitudinal excitation at $\omega_\ast$. What we show below is that the Coulomb interaction lifts the longitudinal/transverse degeneracy and elevates the longitudinal gap frequency from $\omega_\ast$ (while leaving the transverse excitation frequency at $\omega_\ast$) to generate a second massive excitation, the high frequency (first) plasmon at $\omega(k \rightarrow 0) = \omega_+$.    
 
To study the issue we consider a model of a strongly correlated binary Coulomb liquid, consisting of two kinds of, say, positively charged particles of charges $Z_1 e$ and $Z_2 e$, masses $m_1$ and $m_2$, and concentrations $c \equiv c_1$ and $c_2$, respectively; $c \ge 1/2$ can be postulated without loss of generality (BIM -- Binary Ionic Mixture). The two ionic species are immersed in a rigid, neutralizing background. A great deal of work has already been devoted to investigating the equilibrium properties \cite{36,*37,38}, transport coefficients \cite{39,*39b,*39c}, etc. of such a system. There have been also various attempts at understanding correlation induced features in the plasmon dispersion \cite{29,40,19}. Here we present a full analysis of the collective excitations of the system at $k=0$. Our main interest lies in the liquid state, but we will extend our study to the crystalline solid phase as well, primarily with the goal of establishing the link between the excitation spectra in the liquid and solid phases. For the purpose of theoretical analysis we follow the Quasilocalized Charge Approximation (QLCA) \cite{22}. At the same time, in order to  verify the predictions of the theoretical analysis we study the system by detailed molecular dynamics (MD) simulations. Our MD code is an extended version of our earlier code developed for the simulation of one-component (Coulomb) plasmas, based on the PPPM method \cite{30}, allowing particles to carry different charges and to have different masses. For the simulation of the BIM, we use $N=10,000$ particles (or slight different for the simulation of bcc and fcc lattices). For liquid phase conditions the initial positions of the particles are set randomly, for solid phase simulations particles are set at lattice sites. In the measurement phase of the simulation data are collected for the three pair correlation functions, as well as for the quantities needed for the derivation of the dynamical spectra: the spatial Fourier components of the microscopic density and current fluctuations. These are acquired separately for the two species of the binary mixture. A subsequent Fourier transformation in the time domain \cite{10} yields the dynamical structure functions $S_{AB}(k,\omega)$, as well as the longitudinal and transverse current fluctuation spectra, $L_{AB}(k,\omega)$ and $T_{AB}(k,\omega)$, respectively. Collective modes are identified as peaks appearing in these spectra.

The Dynamical Matrix $C$, calculated in the QLCA becomes
\begin{eqnarray}
C^{\mu \nu}_{AB}(\textbf{k}) &=& - \frac{1}{4\pi} \int d^3 \overline{r} \{ \omega_{AB}^2 \psi^{\mu \nu}
(\overline{\textbf{r}}) {\rm e}^{-i \textbf{k}\cdot \textbf{r}} [1 + h_{AB}(r) ] \nonumber \\
&&- \delta_{AB} \sum_{C(all)} \Omega_{AC}^2  \psi^{\mu \nu} (\overline{\textbf{r}}) [1 + h_{AC}(r)] \}  \nonumber \\
&&+ \delta_{AB} \delta^{\mu \nu} \sum_{C(all)} \frac{1}{3} \Omega_{AC}^2 \nonumber \\
\psi^{\mu \nu} (\overline{\textbf{r}}) &=& \frac{1}{\overline{r}^3} \Bigl( 3 \frac{r^\mu r^\nu}{r^2}-\delta^{\mu \nu} \Bigr) 
- \frac{4 \pi}{3} \delta^{\mu \nu} \delta(\overline{\textbf{r}})~~~
\end{eqnarray}
The indices $A,B$ designate the species; $h_{AB}(r)$ is the pair correlation function between particles in species A and B.
\begin{eqnarray}
\omega_{AB}^2 &=& \frac{4 \pi e^2 Z_A Z_B \sqrt{n_A n_B}}{\sqrt{m_A m_B}} \nonumber \\
\Omega_{AB}^2 &=& \frac{4 \pi e^2 Z_A Z_B n_B}{m_A}~~~~
\end{eqnarray}
are the nominal plasma and Einstein frequencies, respectively, in terms of which the elements of the $C$-matrix at $k=0$ become
\begin{eqnarray}
C_{11}^{L}(0) &=& \omega_{11}^2 + \frac{1}{3}\Omega_{12}^2, ~~~~~ C_{22}^{L}(0) = \omega_{22}^2 + \frac{1}{3}\Omega_{21}^2\\
C_{12}^{L}(0) &=& \frac{2}{3}\omega_{12}^2 = \frac{2}{3}\Omega_{12}\Omega_{21}, \nonumber \end{eqnarray}
and
\begin{eqnarray}
C_{11}^{T}(0) &=& \frac{1}{3}\Omega_{12}^2, ~~~~~ C_{22}^{T}(0) = \frac{1}{3}\Omega_{21}^2\\
C_{12}^{T}(0) &=& - \frac{1}{3}\Omega_{12}\Omega_{21}, \nonumber 
\end{eqnarray}
The $L$ and $T$ superscripts designate longitudinal and transverse elements. The collective modes are obtained as the roots of the characteristic equation $||\mathbf{C} - \omega^2\mathbf{I}|| = 0$.

We introduce now the asymmetry parameters $p$ and $q$ with $p^2 = Z_2 n_2 / Z_1 n_1$, and $q^2 = Z_2 m_1 / Z_1 m_2$. In the sequel we express all frequencies in the unit of $\omega_1 \equiv \omega_{11} = \sqrt{4 \pi Z_1^2 e^2 n_1 / m_1}$, and use the notations $Z=Z_2/Z_1$ and $m=m_2/m_1$ for the charge and mass ratios, respectively. Now the resulting gap frequencies become:
\begin{eqnarray}
\omega_\pm^2 &=& \frac{1}{2} ( B \pm \sqrt{\Delta} ) \\
\omega_T^2 &\equiv& \omega_\ast^2 =  \frac{1}{3}(p^2 + q^2) \nonumber  \\
B &=& 1 + p^2 q^2 + \frac{1}{3}(p^2 + q^2) \nonumber \\
\Delta &=& B^2 - \frac{4}{3} q^2 (1 + p^2)^2 \nonumber. 
\end{eqnarray}
$\omega_+$ and $\omega_-$ are the longitudinal plasmons, while $\omega_T$ is a doubly degenerate transverse mode. These results were already anticipated in \cite{21,22}. Note that the gap frequencies                                                                                                                                                                                                                                                                                                                                                                                                                                                                                                                                                                              are ordered as $\omega_+ > \omega_T > \omega_-$. They can also been shown to satisfy a generalized Kohn sum rule \cite{23}, which, however, does not protect them from a rather complex dependence on the system parameters. 

\begin{figure}[htb]
\includegraphics[width=0.6\columnwidth]{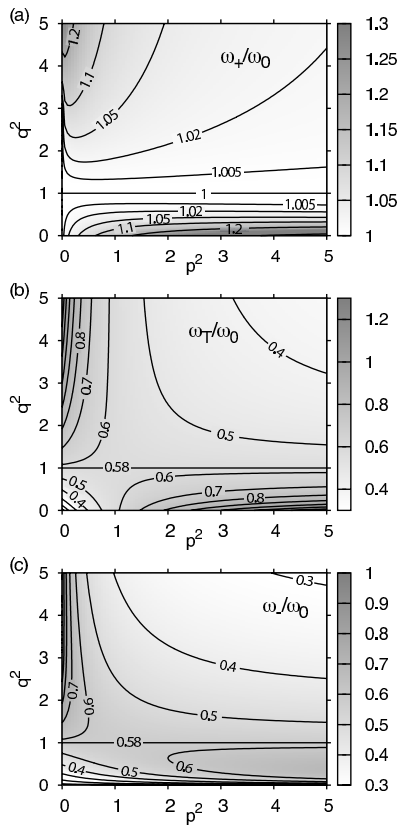}
\caption{\label{fig:fig1}
(color online) Gap frequencies in relation to the RPA plasma frequency $\omega_0 = \omega_{11}\sqrt{1+p^2 q^2}$.}
\end{figure}

In the weak coupling (RPA) approximation there would exist only one excitation frequency $\omega_0 = \omega_{11}\sqrt{1 + p^2 q^2}$. That correlations make $\omega_+ > \omega_0$ has already been noted by Hansen et al. \cite{27}. It has also been suggested \cite{29} that in the strong coupling situation the system develops a hydrodynamic (or virtual average atom) frequency $\bar{\omega} = \sqrt{q^2 / (1+q^2)}(1+p^2)$. While this frequency plays a role in the low frequency acoustic spectrum of the system (similarly to Yukawa systems \cite{20}), it is not part of the spectrum displayed above. In Figure 1 we portray the three predicted gap frequencies in relation to the RPA plasma frequency $\omega_0$, showing the remarkable differences brought about by the strong coupling. The special role of  the $q=1$ structure is visible as an $\omega_+ = 1$, $\omega_- = \omega_T = 1/\sqrt{3}$, $p$-independent separatrix, representing a quasi one-component behavior \cite{19,20,22,27,29,31}.

In the following sequence of graphs we  compare the QLCA predictions with the actual behavior of the system, based on a series of MD simulations over a range of coupling values. We characterize the strength of the overall coupling by the value of $\Gamma \equiv \Gamma_1 = Z_1 e^2 / a_1 k_\text{B} T$, where $a_A^3 = 3/(4\pi n_A)$. Obviously, depending on $Z_2$ and $c_2$, the actual coupling strength can be quite different. A fair measure of its value can be gleaned by observing where the freezing of the liquid sets on. (A more sophisticated determination could be obtained via the linear superposition rule \cite{11}, but is not necessary for our purpose). We have found that defining $\Gamma_\text{eff} = \langle Z \rangle^2 e^2/(a_0 k_\text{B} T_0) = c^{5/3}(1+p^2)^2 \Gamma$ provides a reasonably uniform liquid/solid phase boundary at $\Gamma_\text{eff} \approx 174$, where $\langle Z \rangle = (Z_1n_1+Z_2n_2)/(n_1+n_2)$. Our $\Gamma$ values range from weak/moderate coupling ($\Gamma=1$) moving up into and beyond the crystallization regime ($\Gamma > 150$). As to the lattice structure in the solid phase, we can make predictions only in simple cases: for $c=0.5$  we expect the lattice structure to be bcc, for $c=0.75$ to be fcc. The stability of these lattices at zero temperature has been tested: for the bcc $0.278 < Z < 3.596$, for the fcc $0.731<Z<1.512$.

\begin{figure}[htb]
\includegraphics[width=0.8\columnwidth]{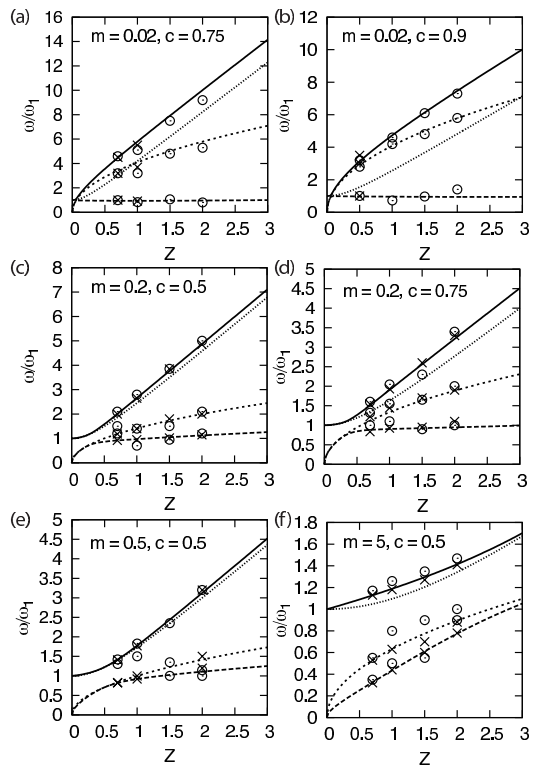}
\caption{\label{fig:fig2}
(color online) Predicted $Z$ and $m$ dependences of the gap frequencies (lines) together with MD results for solid (crosses) and strongly coupled liquid (circles) cases. System parameters are indicated in the figures.}
\end{figure}

In Fig.~2 the predicted $Z$ and $m$ dependences of the gap frequencies are shown for different concentrations along with MD simulation results for high $\Gamma$ values. Shown are also the matching of the liquid gap frequencies with the corresponding values in the crystalline solid. In the bcc there is a one-to-one agreement between the liquid and solid gap frequencies; in the fcc there are additional optic modes, due to the increased number of particles inside the unit cell (cf. \cite{20}). The overall agreement with the theoretical prediction is very good: the analytic description of the mode structure seems to be well confirmed.

It should be emphasized that the QLCA gap frequencies are formally $\Gamma$ independent and the detailed structures of the correlation functions does not enter Eqs.~(6). (This is not true for the $k \neq 0$ behavior, not shown here). The only feature that has been exploited is that $g_{AB}(r=0)=0$; nevertheless the strong coupling approximation is inherent in the model, because the QLCA is built on the localization assumption, a hallmark of the strong coupling. 

\begin{figure}[h!]
\includegraphics[width=\columnwidth]{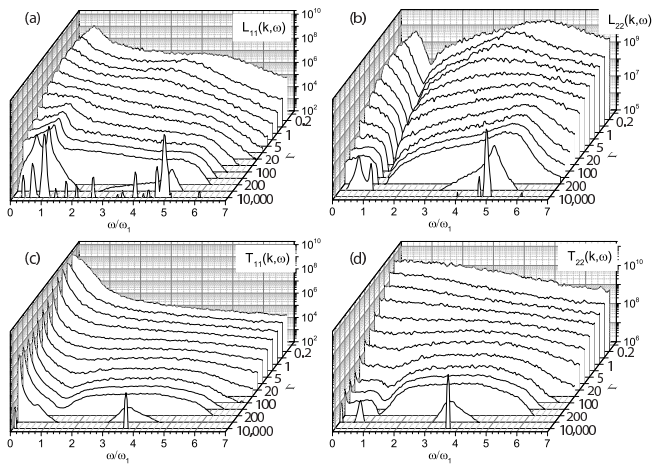}
\caption{\label{fig:fig3}
(color online) Dynamical partial structure functions for a series of coupling parameters $\Gamma$ for $Z=0.8$, $m=0.02$, $c=0.75$ and the lowest wavenumber accessible in the simulation: $ka=0.188$, where $a=\sqrt{a_1a_2}$. The two spectra in the front represent solid (fcc) systems with $\Gamma=10,000$ and 300, respectively.}
\end{figure}

\begin{figure}[h!]
\includegraphics[width=0.7\columnwidth]{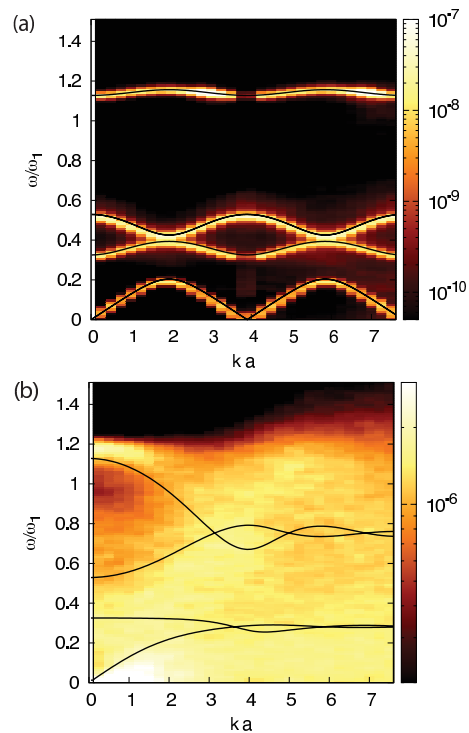}
\caption{\label{fig:fig4}
(color online) Color maps in the background are the sums of current fluctuation spectra, $L_{11}+L_{22}+T_{11}+T_{22}$, from MD simulations for $Z=0.7$, $m=5$, $c=0.5$, and (a) $\Gamma=10,000$ (bcc lattice) and (b) $\Gamma=100$ (liquid). The black lines overlaid are dispersions relations computed by (a) lattice summation, and (b) QLCA for the same system.}
\end{figure}

\begin{figure}[h!]
\includegraphics[width=0.8\columnwidth]{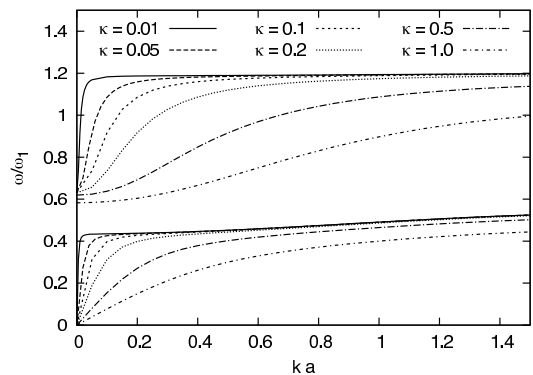}
\caption{\label{fig:fig5}
(color online) Longitudinal dispersion curves calculated for binary Yukawa bcc lattices for a series of low Yukawa screenings (characterized by $\kappa = a/\lambda_D$) and $Z=1$, $m=5$.}
\end{figure}

More insight into how strong coupling affects the dynamics of the system can be gleaned from Fig.~3, where we show the dynamical partial structure functions $L_{11}(\omega)$, $L_{22}(\omega)$, $T_{11}(\omega)$, $T_{22}(\omega)$, $L(\omega)$ representing the spectra of the longitudinal, $T(\omega)$ of the transverse current fluctuations. We present sequences of graphs for a selected set of parameters  where the effect of the increasing strength of the coupling can be followed, from low coupling ($\Gamma=0.2$) into well in the crystalline phase. We observe that at low $\Gamma$ values the system exhibits RPA behavior, where only one gapped mode $\omega_0$ survives. The characteristic strong coupling behavior with the appearance of the $\omega_+$ first plasmon and the  $\omega_-$ second plasmon takes place around $\Gamma=40$, while the transverse modes appear later, around $\Gamma=100$. (Note, however, that these $\Gamma$ values are appropriate for the chosen system parameters only, and they vary with the change of the system parameters). There seems to be a ``no-man's land'', somewhere between $\Gamma=10$ and $\Gamma=40$, where virtually no collective excitation exists. A remarkable feature can be observed in the $L_{22}(\omega)$ (2 is the light component) graph: the development of a very well defined Fano-like sharp minimum, at an $\omega$ value adjacent to $\omega_-$ ; a somewhat similar feature, a sharp drop, appears in $L_{11}(\omega)$ (1 is the heavy component), adjacent to $\omega_+$. A discussion and attempted explanation of these unexpected features will be presented elsewhere.

While the focus of this Letter is a presentation and discussion of the $k=0$ gapped excitations, it is instructive to examine a sample of the full $\omega(k)$ dispersions. This is done in Fig.~4, where for two representative sets of parameters the MD results are portrayed, both for the liquid and for the crystalline solid phase, accompanied by theoretical dispersion curves calculated with the aid of the QLCA and the harmonic phonon approximation, respectively. The most important feature to observe is the appearance of the acoustic doubly degenerate transverse mode with a sound speed $s$ of the order of $s \sim \bar{\omega} a$. More detailed discussion of the dispersion is also deferred to a later presentation.

Finally, it is instructive to explore the details of the transition from a well behaved Yukawa system with $\kappa \neq 0$ to the singular Coulomb ($\kappa=0$) case. A sequence of dispersion curves illustrating the process as described in the introductory paragraph                                                                                                                                                      is given in Fig.~5. While the transition is discontinuous at $k=0$, it becomes quasi-continuous in the $k > \kappa$ domain.	

In summary, we have shown that in the strongly coupled phase of a binary system of charged particles the excitation spectrum of collective modes dramatically changes from the simple structure that exists in the domain of weak coupling: the single plasmon with the combined plasma frequency $\omega_0$ of the two species $\omega_0 = \sqrt{\omega_{10}^2+\omega_{20}^2}$ is replaced by the doublet of a new type of excitations, a high frequency first plasmon $\omega_+$ and a low frequency, second plasmon $\omega_-$. This second plasmon is generated by the Anderson mechanism from the longitudinal acoustic Goldstone boson that one would have in a system with a short range interaction. The first plasmon has a frequency always higher, but (depending on the system parameters) most of the time not much higher than $\omega_0$. There is, however, no smooth transition from the weak coupling $\omega_0$ to the strong coupling $\omega_+$. In the intermediate coupling domain no collective excitation can be detected.    

As to the remaining part of the spectrum, we have confirmed the existence of transverse excitations, consisting of a set of doubly degenerate acoustic and a set of doubly degenerate gapped modes. The acoustic speed is governed by the oscillation frequency of the ``virtual atom'' \cite{20} (hydrodynamic frequency \cite{29}).

The theoretically predicted spectrum has been verified by detailed MD computer simulations and for the $n_1=n_2$ case the matching of the liquid dispersion and the calculated bcc lattice dispersion at $k=0$ has been demonstrated.

The physical systems that come closest to the realization of the simplified model investigated in this paper are brown dwarf interiors \cite{41,*42}, carbon-oxygen stars in their helium shell burning phase \cite{32,*33}, and trapped Be$^+$ - Xe$^{+44}$ ionic mixtures realized in the Electron Beam Ion Trap program at Lawrence Livermore National Laboratory \cite{38}, as well as mixtures of fermion gases, such as constituted by electrons in transition metals \cite{44} and in heavy fermion systems \cite{45}.

\begin{acknowledgments}
This work has been supported by the NSF Grants 0715227, 0813153, 1105005, 0812956, OTKA Grants K-105476 and NN-103150. 
\end{acknowledgments}

%

\end{document}